# Proposal of the Readout Electronics for the WCDA in LHAASO Experiment[*]

ZHAO Lei(赵雷)[1,2] LIU Shu-Bin(刘树彬)[1,2] AN Qi(安琪)[1,2;1]

[1]State Key Laboratory of Particle Detection and Electronics, University of Science and Technology of China, Hefei, 230026, China,
[2]Department of Modern Physics, University of Science and Technology of China, Hefei, 230026, China

**Abstract:** The LHAASO (Large High Altitude Air Shower Observatory) experiment is proposed for very high energy gamma ray source survey, in which the WCDA (Water Cherenkov Detector Array) is the one of the major components. In the WCDA, a total of 3600 PMTs are placed under water in four ponds, each with a size of 150 m × 150 m. Precise time and charge measurement is required for the PMT signals, over a large signal amplitude range from single P.E. (photo electron) to 4000 P.E. To fulfill the high requirement of signal measurement in so many front end nodes scattered in a large area, special techniques are developed, such as multiple gain readout, hybrid transmission of clocks, commands, and data, precise clock phase alignment, and new trigger electronics. We present the readout electronics architecture for the WCDA and several prototype modules, which are now under test in the laboratory.

## 1 Introduction

Gamma ray source detection above 30 TeV is an encouraging approach for finding galactic cosmic ray sources. All sky survey for gamma ray sources is essential for population accumulation for various types of sources above 100 GeV [1]. To target these goals, the Large High Altitude Air Shower Observatory (LHAASO) [1, 2] is proposed, with multiple air shower detection techniques combined together. As shown in Fig. 1, the LHAASO mainly consists of a KM2A (1 km$^2$ complex array), a WCDA (Water Cherenkov Detector Array) [3, 4], a WFCTA (wide FOV Cherenkov telescope array) and a SCDA (high threshold core-detector array).

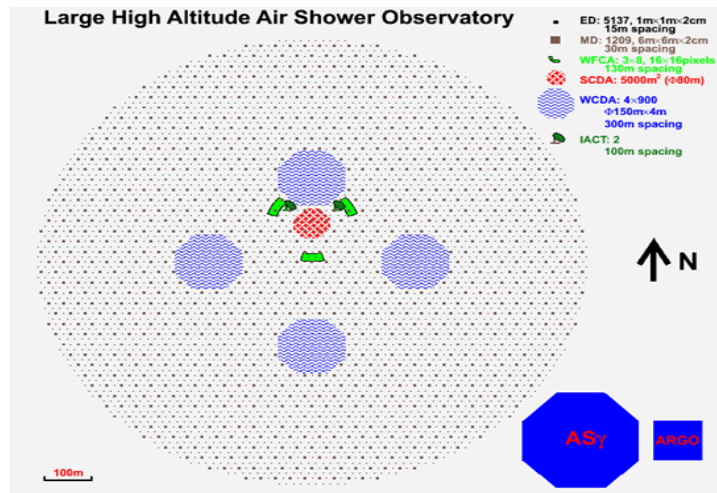

Fig. 1. Layout of the LHAASO array.

[*]Supported by the Knowledge Innovation Program of the Chinese Academy of Sciences (KJCX2-YW-N27) and the National Natural Science Foundation of China (11175174, 11005107)
1) Email: anqi@ustc.edu.cn

As one of the major components in LHAASO, the WCDA consists of four 150 m × 150 m ponds, as marked by the four octagons in Fig. 1, and thus a total area of 90000 m$^2$ is covered. Every pond incorporates 900 detector cells (5 m × 5 m), each with one photomultiplier tube(PMT) looking up at the bottom center to collect the Cherenkov light produced by the shower particles in water. This detector array of 900 cells is partitioned into 10 × 10 detector units, as shown in Fig. 1. Each unit contains 3 × 3 PMTs and their outputs are fed to one front end electronics module (FEE) for charge and time information measurement.

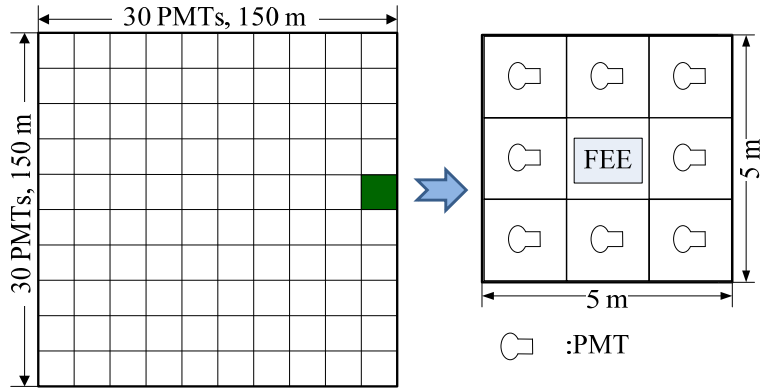

Fig. 1. Structure of one WCDA pond in LHAASO.

Both precise time and charge measurement is required in the WCDA readout electronics. We need to read out a total of 3600 PMT channels scattered within a large area, and the RMS of time measurement is required to be better than 0.5 ns, which means precise clock distribution and phase alignment in a large scale is paramount. Besides, data, commands, and clocks need to be transferred over a long distance (hundreds of meters), so the transmission quality and the system complexity are great concerns. To reduce the number of signal cables, we decide to propose a new architecture, instead of employing the traditional trigger structure that requires receiving hit information from front ends and feeding trigger signals back. Furthermore, the large input dynamic range (1~ 4000 P.E.) is also a great challenge, compared with the readout electronics for the WCDAs in other experiments [5, 6].

The measurement requirement of the readout electronics is listed in Table. 1.

Table 1. Measurement requirement of the WCDA readout electronics.

| Item | Requirement |
| --- | --- |
| Bin size of time measurement | < 1 ns |
| RMS of time measurement | < 0.5 ns |
| Dynamic range of time measurement | 2 us |
| Minimum interval of two hit signals | 100 ns |
| Dynamic range of charge measurement | S.P.E. ~ 4000 P.E. |
| Resolution of charge measurement | 30%@S.P.E.;    3%@4000P.E. |
| Channel number | 3600 |

Considering the situation in the read out of the WCDA in LHAASO, the following difficulties needs to be overcome with novel methods and techniques:

1) precise time and charge measurement over an ultra large dynamic range;
2) high-quality clock distribution and phase alignment;
3) long distance transmission of clocks, commands, and huge amount of data;
4) new trigger electronics design.

## 2 Architecture of the WCDA Readout Electronics

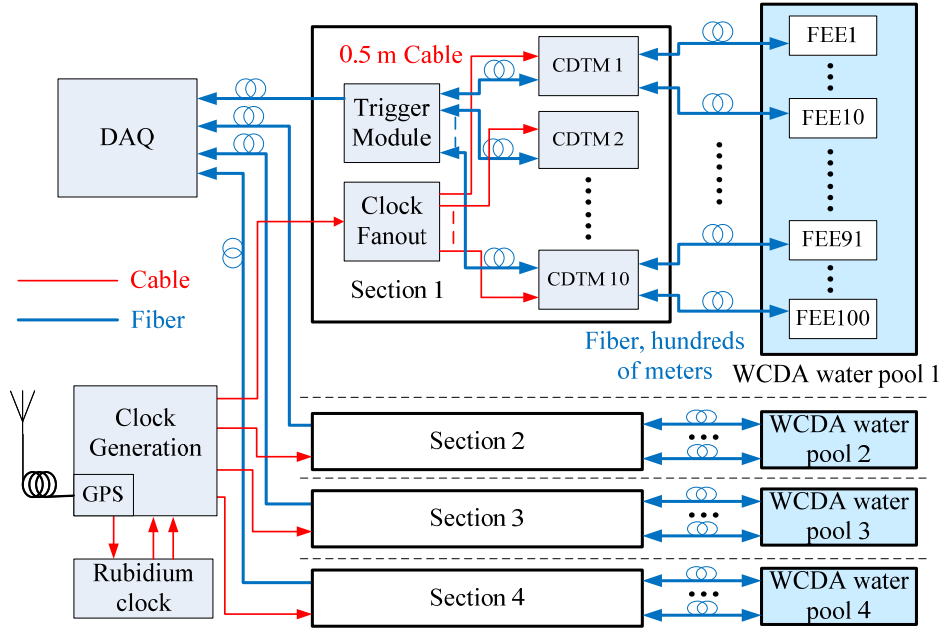

Fig. 2. Architecture of the WCDA readout electronics.

The architecture of the WCDA readout electronics is shown in Fig. 2. For each WCDA pond, 100 front end electronics modules are located above water, very close to the PMTs. Each FEE is responsible for the signal readout for 9 PMTs within one detector unit in Fig. 1. The signal from the PMT is directly digitized by the FEE, and then transmitted to the clock and data transferring module (CDTM) in the control room. Signal transmission is based on fibers (hundreds of meter long) to guarantee a good transmission quality. Special methods are applied to address the issue of large dynamic range in FEEs, which will be presented in posterior sections.

Each CDTM collects data from 10 FEEs and then transfers the data to the trigger module; it also functions as the bridge for communication between the FEEs and the control system for commands transferring as well as system monitoring. Due to the complicated phenomena of cosmic ray showers, a high flexibility of trigger electronics is required to accommodate different potential trigger patterns. Invoked by the "triggerless" idea, we propose a new trigger electronics structure to achieve a good flexibility.

As mentioned above, a high time measurement resolution better than 1 ns is required over all the 3600 FEE channels. Therefore, we need to achieve a high quality clock distribution. Based on a simplified White Rabbit (WR) protocol [7-9], we employed an adaptive phase adjustment method to precisely align the clock phases. Besides, absolute time information needs to be included in the measurement results to identify an air shower event. We receive the signal from the global positioning system (GPS), and use it to discipline a rubidium clock source to generate the system clock reference, as shown in Fig. 2. Detailed information is included in following sections.

### 2.1 Measurement method of the PMT signal with an ultra large dynamic range

The PMT signal amplitude would vary from 1 P.E. to 4000 P.E. while the discrimination threshold is as low as 0.25 P.E. It means that a resolution of 16 bits is required on the overall system performance, which is very difficult to achieve. To guarantee a good measurement resolution, especially for small signals (~ 1 P.E.), we employ two readout channels for each PMT: one is for the anode and the other is for a dynode (e.g. Dynode 10), as shown in Fig. 3.

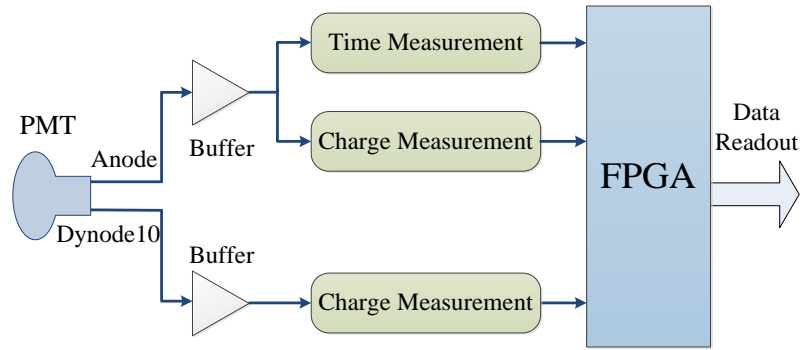

Fig. 3. Readout of the PMT signal.

By adjusting the gains of the buffers in the two readout channels, the anode channel can effectively cover a range of 1~ 133 P.E. while the Dynode 10 channel covers a range of 30 ~ 4000 P.E. A total dynamic range of 4000 is achieved, and there exists enough overlap between the two channels. Since the dynamic range has been reassigned to two smaller scales, the requirement on the electronics is greatly reduced. The PMT signal from the anode is also used for time measurement. After passing through the discrimination circuits, the hit signal is digitized by a Time to Digital Converter (TDC) integrated in the FPGA.

As for precise charge measurement, we can employ the digital peaking method, in which the PMT is digitized by an ADC after amplification and shaping, and then the peak value of the amplitude can be sought based on the digital signal processing. We have worked on the measurement scheme [10-13], and finished the design of a FEE prototype base on this method, as shown in Fig. 4. The performance tests in the laboratory are underway.

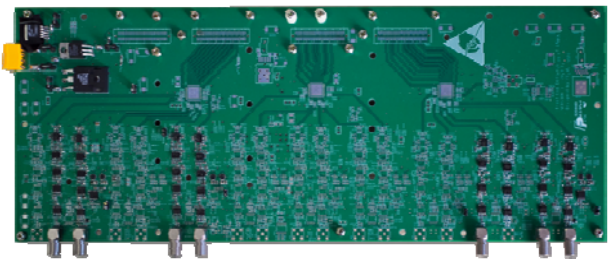

Fig. 3. Photograph of the FEE prototype.

Besides, we are also considering applying the Time-Over-Threshold (TOT) method in the signal measurement, and integrating the circuits within an ASIC (application specific integrated circuit ) chip. The ASIC design is also in process.

**2.2 High quality clock distribution over a large area**

Since the FEEs are distributed over a large area, the delay difference among long signal transmission paths would cause non-aligned phases between the clock signals of different FEEs, which cannot be tolerated in high resolution time measurement. Besides, aging of the electronics and temperature variation would lead to fluctuation of the time delay. Thus, a precise real-time delay calibration and adjustment is needed.

A novel technology named White Rabbit [7-9] is recently developed in the frame of CERN's (and GSI's) renovation projects. It is capable of controlling thousands of nodes over a few kilo-meters with a good timing accuracy. As for the timing, its kernel technique is based on the Precision Time Protocol (PTP, IEEE1588) and the Synchronous Ethernet technology. Invoked by the WR technique, we employ an adaptive delay measurement and compensation structure, as shown in Fig. 4.

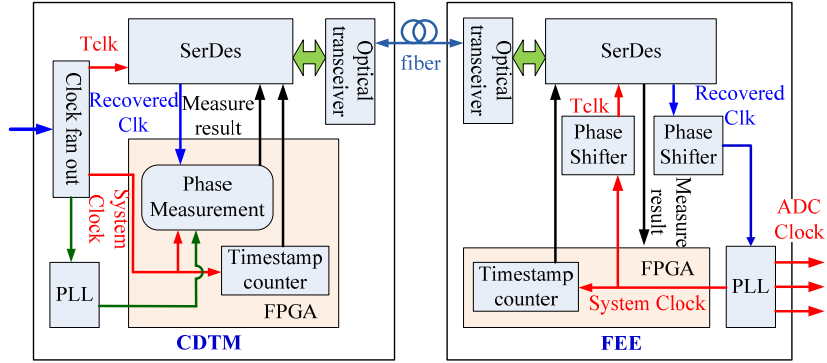

Fig. 4. Block diagram of the clock distribution and delay compensation method.

As aforementioned, each CDTM communicates with 10 FEEs over a distance of hundreds of meters. The CDTM (as the master) distributes the clock signals to the FEEs (as the slaves). Real time phase measurement is implemented in an FPGA based on the digital dual mixer time difference [14] (DMTD) measuring method. To further enhance the measurement resolution, an FPGA based TDC (resolution ~ 1 ns) is used. The phase measurement results are sent to the FEEs, in which the clock signal is adjusted for phase alignment. Digitally controlled phase locked loop (PLL) and precise delay line ASICs are used to achieve a tuning step of around 10 ps.

As a high-quality timing system, all the FEEs are required to be initiated and started at a precisely matched time point, which means a global reset must arrive at all the FEEs simultaneously. This is done with two steps. First, the clock signals are well aligned using the method just mentioned above. Second, a special timing sequence is designed, as shown in Fig. 5. The master (CDTM) sends a 'Reset' pulse just after the PPS (pulse per second) signal. When this reset signal is detected by the slave (FEE), a flag named 'Reset ready' is asserted, with which the local reset signal in the slave can be generated after the next PPS signal is detected. To guarantee a synchronous reset of all the slaves, it is required that the uncertainty of the PPS signal is within one system clock cycle, while the uncertainty of the reset signal from the master can be as large as 1 s, which can be easily achieved.

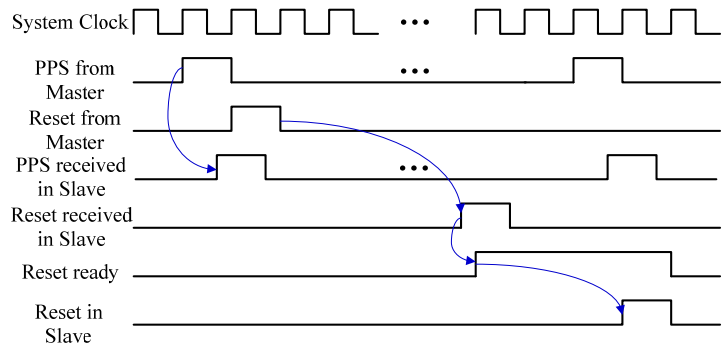

Fig. 5. Timing sequence for a global reset.

Besides, the CDTM is also responsible for collecting data from the FEEs, and transferring them to the trigger processing electronics. Considering an average raw event rate of 50 kHz, the total data rate for each CDTM would be as high as 50 kHz × 96 bit (packet size) × 9 (FEE channel number) × 10 = 432 Mbps. Considering the cost of 8 b/10 b encoding [15], the data rate is around 540 Mbps. Fibers are employed to achieve such a high transmission speed. Meanwhile, data, commands, and clocks are mixed together in the fiber based transmission line, so a good system simplicity is guaranteed.

## 2.3 New trigger electronics design

In traditional trigger electronics such as in AMANDA [16] and in BES III [17], hit signals are collected from FEEs for trigger processing, and global triggers need to be sent back to front ends for data readout. The complex phenomena of cosmic ray showers require a highly flexible trigger electronics, which can be modified online to accommodate different potential trigger patterns.

Invoked by the "triggerless" idea [18, 19], we employ a new structure of trigger electronics. Complete data packets are collected from all the FEEs for trigger processing, so much more detailed information can be utilized to achieve a better trigger flexibility. Meanwhile, it also eliminates the need of feeding trigger signals back to front ends; thus a better system simplicity can be achieved. We plan to implement the triggerless idea with two potential methods. The first one is complete triggerless design. All the data from the FEEs are processed by the software in the DAQ. A best flexibility can be guaranteed by this method, of course accompanied by the requirement of more resource cost on data storage and processing. In the second method, we consider designing a trigger module to collect all the data form FEEs, and conducting real-time trigger processing with in an FPGA device. The recent development of data transmission and FPGA techniques makes it a feasible way. The abundant programmable logic resources in the FPGA lead to a reconfigurable trigger electronics. Besides, since a large partition of fake events are rejected by the trigger electronics, the requirement on the DAQ is also reduced.

We have finished the design of a trigger electronics prototype based on the second idea [20], and evaluate this design with a basic trigger pattern. As shown in Fig. 6, the PMTs within one pond are divided into 16 overlapped trigger clusters (each cluster is a $12 \times 12$ PMT array). When more than 12 PMTs in any cluster are fired within any 250 ns duration, it will be recognized as a valid event and a global trigger signal will be generated.

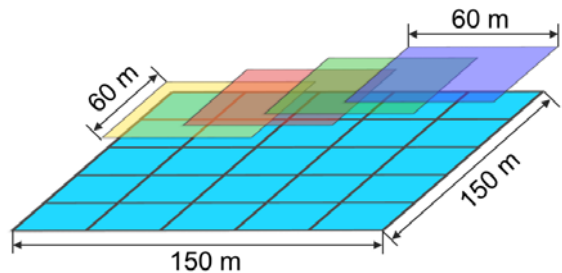

Fig. 6. Cluster arrangement for a basic trigger pattern..

We implement the above trigger pattern with the logic design in one FPGA (Xilinx Virtex-6 xc6vlx240t), as shown in Fig. 7. Trigger processing and valid data selection are both integrated in the FPGA. One trigger module is required for each pond, which means the data from 100 FEEs are accumulated through 10 CDTMs to this module, rendering a raw data rate up to 5.4 Gsps. Fibers are also used here to guarantee a high data transfer rate.

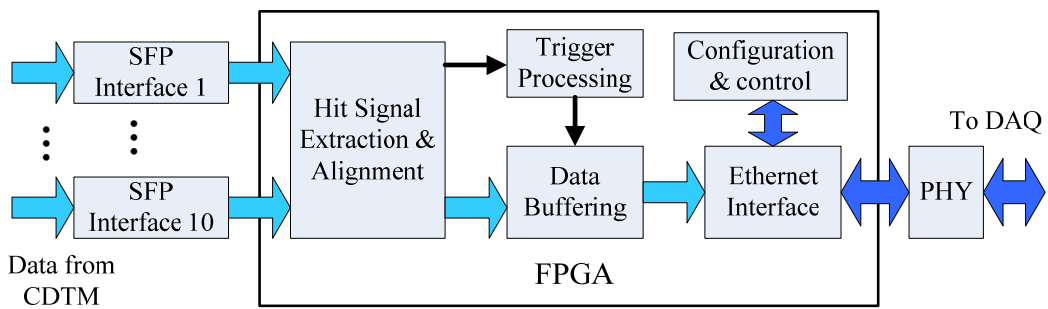

Fig. 7. Block diagram of the LHAASO trigger module.

Tests have been conducted on the data transmission performance and the validity of the trigger processing. As shown in Fig. 8 is the system under test.

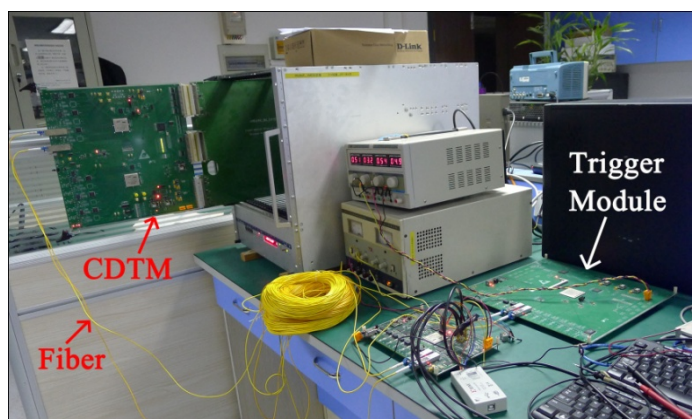

Fig. 8. System under test.

The aforementioned complete triggerless architecture is also under consideration.

## 3 Summary

The requirement and challenge of the WCDA readout electronics in LHAASO are introduced. To overcome the difficulties, we propose special techniques, such as multiple gain readout, hybrid transmission of clocks and commands & data, precise clock phase alignment, and new trigger electronics. The architecture design of the electronics is presented in this paper. Technical design is in process, and we have finished the design of several electronics prototype, which are now under tests in the laboratory.

**Acknowledgement**

This work is supported by the Knowledge Innovation Program of the Chinese Academy of Sciences (KJCX2-YW-N27) and the National Natural Science Foundation of China (No. 11175174 and No.11005107). The authors would like to appreciate Dr. CAO Zhen, HE Hui-hai, YAO Zhi-guo, CHEN Ming-jun, LI Cheng, and TANG Ze-bo for their kindly help. And last, the authors thank all of the LHAASO collaborators who helped to make this work possible.

**References**

1 CAO Zhen, A future project at tibet: the large high altitude air shower observatory (LHAASO), Chin., Phys. C (HEP & NP), 2010, 34(2): 249-252
2 HE Hui-hai et al. LHAASO Project: detector design and prototype, Proceedings of the 31st ICRC, 2009
3 CHEN Ming-jun et al. R&D of LHAASO-WCDA, Proceedings of the 32nd ICRC, 2011
4 YAO Zhi-guo et al. Design & Performance of LHAASO-WCDA Experiment, Proceedings of the 32nd ICRC, 2011
5 The LBNE Collaboration, LBNE Conceptual Design Report, Sept. 17th, 2010
6 Zheng Wang et al. PMT FEE Board Technical Design Report, 2nd Draft, Jun3 18th, 2008
7 Guanghua Gong et al. Sub-nanosecond Timing System Design and Development for LHAASO Project, Proceedings of ICALEPCS2011, 2011
8 J.Serrano, The White Rabbit Project, Proceedings of ICALEPS2009, 2009
9 Tomasz Wlostowski, Precise time and frequency transfer in a White Rabbit network, master of science thesis, Warsaw University of Technology. 2011
10 HAO Xinjun et al. A digitalizing board for the prototype array of LHAASO WCDA, Nuclear Science and Techniques, 2011, 22(3): 178-184
11 CAO Zhe et al. Evaluation Design of the Distributed Electronics System for the Water Cherenkov Detector Arrays in the LHAASO, Nuclear Electronics& Detection Technology, 32(3):243-247
12 HAO Xinjun, The Research on the Readout Electronics for the Prototype Array of LHAASO WCDA, PhD. Thesis, Hefei: University of Science and Technology of China, 2011
13 CAO Zhe, The Research on Distributed Electronics of Water Cherenkov Detector Arrays of LHAASO in Yangbajing of Tibet, PhD. Thesis, Hefei: University of Science and Technology of China, 2011
14 G.Brida, High resolution frequency stability measurement system, Review of Scientific instruments, 2002, 73(5): 2171-2174
15 Widmer A X et al. IBM Journal of Research and Development, 1983, 27: 440-451
16 Andres E et al. Astroparticle Physics, 2000, 13(1):1-20
17 LIU Zhen-an et al. Trigger System of BESIII, Proceedings of the 15th Real-Time Conference, 2007
18 Essel H G et al. Future DAQ for CBM: On-line Event Selection, IEEE Trans. Nucl. Sci., 2006, 53(3): 677-681

19 Appelbe E et al. The GREAT Triggerless Total Data Readout Method, IEEE Trans. Nucl. Sci., 2001, 48(3): 567-569
20 YAO Lin et al. A Prototype of Trigger Electronics for LAWCA Experiment, Chin., Phys. C (HEP & NP), accepted.